\documentstyle[prd,aps,amsfonts,preprint,tighten]{revtex}
\begin{document}
\preprint{\vbox to 50 pt{\hbox{IHES/P/93/48}\hbox{CPT-93/P.2941}}}
\draft
\title{Testing for Preferred-Frame Effects in Gravity\\
with Artificial Earth Satellites}
\author{Thibault Damour}
\address{Institut des Hautes Etudes Scientifiques, 91440 Bures sur
Yvette, France\\
and D\'epartement d'Astrophysique Relativiste et de Cosmologie,
Observatoire de Paris,\\
Centre National de la Recherche Scientifique, 92195 Meudon, France}
\author{Gilles Esposito-Far\`ese}
\address{Centre de Physique Th\'eorique, Centre National de la
Recherche Scientifique,\\
Luminy, Case 907, 13288 Marseille Cedex 9, France}
\date{October 8, 1993}
\maketitle
\begin{abstract}
As gravity is a long-range force, one might {\it a priori\/} expect
the Universe's global matter distribution to select a preferred rest
frame for local gravitational physics. At the post-Newtonian
approximation, two parameters suffice to describe the phenomenology
of preferred-frame effects. One of them has already been very
tightly constrained ($|\alpha_2|<4\times 10^{-7}$, 90\% C.L.), but
the present bound on the other one is much weaker
($|\alpha_1|<5\times 10^{-4}$, 90\% C.L.). It is pointed out that
the observation of particular orbits of artificial Earth satellites
has the potential of improving the $\alpha_1$ limits by a couple of
orders of magnitude, thanks to the appearance of small
divisors which enhance the corresponding preferred-frame effects.
There is a discrete set of inclinations which lead to arbitrarily
small divisors, while, among zero-inclination (equatorial) orbits,
geostationary ones are near optimal. The main $\alpha_1$-induced
effects are: (i)~a complex secular evolution of the eccentricity
vector of the orbit, describable as the vectorial sum of several
independent rotations; and (ii)~a yearly oscillation in the longitude
of the satellite.
\end{abstract}

\narrowtext
\section{Introduction}
The absence of preferred frames in local experiments (or local
boost invariance) is verified everyday in high-energy experiments
but is much more difficult to test for the gravitational interaction.
Metrically coupled tensor--scalar theories of gravity (\`a la
Jordan--Fierz--Brans--Dicke), including general relativity, do not
predict any violation of this invariance (see {\it e.g.} \cite{DEF1}
and references therein). On the other hand, one expects the Universe's
global matter distribution to select a preferred rest frame for the
gravitational interaction if it is mediated in part by a long-range
vector field (or a second tensor field besides the unique one
postulated by Einstein) \cite{WN72}. Tests of the boost-invariance of
gravity in localized systems are therefore of special interest to
determine the {\it field content\/} of the gravitational
interaction, {\it i.e.} specifically whether gravity contains,
besides the standard tensor contribution and an often considered
scalar one, extra contributions due to the exchange of a vector or
a second tensor field.

In the post-Newtonian limit, all preferred frame effects are
phenomenologically describable by only two parameters, $\alpha_1$
and $\alpha_2$ \cite{WN72}. [Note that the post-Newtonian approach
assumes that all the fields contributing to gravity are massless, or
at least have a range much larger than the size of the considered
$N$-body system.] These two preferred-frame parameters contribute
non-boost-invariant terms in the Lagrangian, depending on the
velocities ${\bf v}^0$ of the bodies with respect to some
gravitationally preferred rest frame. More precisely, the
post-Newtonian Lagrangian describing the interaction between 
$N$ {\it spherical\/} bodies reads
\begin{eqnarray}
L^{N{\rm\ body}} = && L_{\beta, \gamma, \eta} +\, L_{\alpha_1} +
L_{\alpha_2} \ ,  \label{eq:1} \\
L_{\beta,\gamma, \eta} = && \sum_A - m_A c^2
  \left( 1- ({\bf v}^0_A)^2 /c^2 \right)^{1/2} \nonumber \\
  && +\, {1\over 2} \sum_{A\not= B} \, {G_{AB}m_Am_B\over r_{AB}}\,
  \left[ 1 + {1\over 2c^2}\,
  \left( ({\bf v}^0_A)^2 + ({\bf v}^0_B)^2 \right)\,-\,
   {3\over 2c^2}\, ({\bf v}^0_A \cdot {\bf v}^0_B) \right. \nonumber \\
  && \qquad\qquad\qquad\qquad\left. -\, {1\over 2c^2}\, ({\bf
n}_{AB}\cdot {\bf v}^0_A)
   ({\bf n}_{AB} \cdot {\bf v}^0_B)\,+\,
    {\gamma \over c^2}\, ({\bf v}^0_A - {\bf v}^0_B)^2 \right]
\nonumber\\
  && -\, {1\over 2} \sum_{B\not= A\not= C}
    (2\beta -1)\, {G^2m_Am_Bm_C\over c^2 r_{AB}r_{AC}}\ ,
    \label{eq:2} \\
L_{\alpha_1} = && -\,{\alpha_1 \over 4} \sum_{A\not= B}\,
    {G m_Am_B\over c^2 r_{AB}}\, ({\bf v}^0_A \cdot {\bf v}^0_B) \ ,
    \label{eq:3} \\
L_{\alpha_2} = && {\alpha_2\over 4} \sum_{A\not= B}
   {Gm_A m_B \over c^2 r_{AB}}\,
  \left[ ({\bf v}^0_A \cdot {\bf v}^0_B) -
         ({\bf n}_{AB} \cdot {\bf v}^0_A)
         ({\bf n}_{AB} \cdot {\bf v}^0_B)  \right]\ , \label{eq:4}
\end{eqnarray}
where the effective coupling constant $G_{AB}$ for the gravitational
interaction of bodies $A$ and $B$ is
\begin{equation}
G_{AB} = G\left[1+\eta\left({E_A^{\rm grav}\over m_Ac^2} +
{E_B^{\rm grav}\over m_Bc^2}
\right)\right]
\ ,
\label{eq:4bis}
\end{equation}
in which $\eta$ denotes the combination of parameters
\begin{equation}
\eta\equiv 4\beta-\gamma-3-\alpha_1+{2\over 3}\alpha_2
\ ,
\label{eq:4ter}
\end{equation}
while $E_A^{\rm grav}\equiv -(G/2)\int_A\int_Ad^3x d^3x'\rho({\bf
x})\rho({\bf x}')/|{\bf x}-{\bf x}'|$ denotes the gravitational
self-energy of body $A$. Apart from the contributions of $\alpha_1$
and $\alpha_2$ to $G_{AB}$ ({\it via\/} $\eta$),
$L_{\beta,\gamma,\eta}$ is the usual boost-invariant Lagrangian
involving the Eddington parameters $\beta$ and $\gamma$, which are
sufficient to parametrize the whole class of metrically-coupled
tensor--multi-scalar theories ($\beta=\gamma=1$ in general relativity,
while these parameters can take arbitrary values in tensor--scalar
theories). The parameters $\alpha_1$ and $\alpha_2$ contribute to the
effects associated with a violation of the strong equivalence
principle ($G_{AB}\neq G$) only through the combined parameter
$\eta$. Observational tests of the parameter $\eta$ are discussed in
the literature ({\it e.g.} \cite{W81}). In the following we
concentrate on the effects associated to the velocity-dependent
Lagrangian $L_{\alpha_1}$ of Eq. (\ref{eq:3}).

It has been shown in Ref. \cite{N87a} that the close alignment of the
Sun's spin axis with the solar system's planetary angular momentum
yields an extremely tight bound on $\alpha_2$
\begin{equation}
|\alpha_2| < 3.9\times 10^{-7} \qquad
\mbox{(90 \% C.L.)}\ .
\label{eq:5}
\end{equation}
This limit on $\alpha_2$ is much stronger than the existing limits
on the other post-Newtonian parameters $\beta$, $\gamma$ and
$\alpha_1$ \cite{F1}. Present experimental bounds on the Eddington
parameters are at the level \cite{W81}
\begin{equation}
|\gamma-1| < 3\times 10^{-3}\quad,\quad |\beta -1|< 3\times 10^{-3}
\qquad
\mbox{(90 \% C.L.)}\ ,
\label{eq:5bis}
\end{equation}
while the limits on $\alpha_1$ are about six times tighter
\cite{W81,NW72,WG76,H84,DEF2}. More precisely combined orbital data
on the planetary system yield \cite{H84}
\begin{equation}
 \alpha_1 = (2.1 \pm 3.1) \times 10^{-4} \qquad
\mbox{(90 \% C.L.)}\ ,
\label{eq:6}
\end{equation}
while the bound obtained in the strong-field regime by analyzing
binary-pulsar data is comparable \cite{DEF2}
\begin{equation}
|\alpha_1| < 5.0\times 10^{-4} \qquad
\mbox{(90 \% C.L.)}\ .
\label{eq:7}
\end{equation}

Recent theoretical developments in tensor--scalar cosmological models
\cite{DN} suggest that a natural level for $|\beta-1|$ and $|\gamma
-1|$ is $\sim 10^{-5}$---$10^{-7}$. The generalization of this
result to classes of gravitational theories involving extra vector
or tensor interactions has not been worked out, but, by analogy, one
might expect the present limit on $\alpha_1$ to be too weak to
constrain at a significant level the participation to gravity of
extra vector or tensor fields. It is therefore worth discussing
experiments having the capability of improving the existing limits
on $\alpha_1$ down to the $10^{-5}$---$10^{-6}$ level.

The object of the present paper is to show that artificial
Earth satellites offer very promising tools to improve the precision
of measurement of $\alpha_1$ by, possibly, a couple of orders of
magnitude. Indeed, we find that the appearance of small divisors
can considerably enhance preferred-frame effects when the
semi-major axis of the satellite's orbit and/or its inclination with
respect to the Earth's equatorial plane take particular values.
Section II is devoted to the secular evolution equations
satisfied by the orbital elements, whose $\alpha_1$-dependent
contributions derive from the disturbing function (\ref{eq:10}). As a
first example of a small divisor which enhances preferred-frame
effects, we consider in section III the simple case of
equatorial orbits, and show that because of a competition between
tidal forces and the quadrupolar moment of the Earth, there exists an
optimal value of the semi-major axis which maximizes the
perturbations due to $\alpha_1$. It turns out that geostationary
satellites are nearly optimal. The case of non-equatorial orbits
is studied in section IV. We point out that preferred-frame effects
can be enhanced by arbitrarily large factors if the inclination is
sufficiently close to one of six specific values, so that the
accuracy of measurement of $\alpha_1$ becomes limited only by the
finite duration of the experiment.

When deriving bounds on $\alpha_1$ or $\alpha_2$ [such as Eqs.
(\ref{eq:5}),(\ref{eq:6}),(\ref{eq:7})], it is necessary to make a
definite assumption about the preferred rest frame entering the
Lagrangians (\ref{eq:3}),(\ref{eq:4}). The standard assumption 
\cite{W81,N87a,NW72,WG76,H84,DEF2}, that we shall take up in the
present paper, is to choose the frame defined by the cosmic microwave
background [this essentially means that the range of the
putative extra vector or tensor field responsible for
the violation of local boost invariance is at least of
cosmological magnitude]. However, we shall see in section V.A that
(somewhat weaker) bounds on $\alpha_1$ can be obtained
without the need of such a hypothesis. Indeed, the `absolute' velocity
of an artificial Earth satellite can be decomposed as
${\bf v}^0_{\rm sat} = {\bf v}_{\rm sat} + {\bf v}_\oplus + {\bf
w}$, where ${\bf v}_{\rm sat}$ is the velocity of the satellite with
respect to the Earth, ${\bf v}_\oplus$ the orbital velocity of the
Earth around the Sun, and $\bf w$ the velocity of the Sun
with respect to the gravitationally preferred frame. The
$\alpha_1$-contribution (\ref{eq:3}) to the interaction term between
the Earth and the satellite reads then
\begin{equation}
L_{\alpha_1} = -\,{\alpha_1\over 2}\,{Gm_\oplus m_{\rm sat}\over c^2
r_{\oplus{\rm sat}}}\,({\bf v}_\oplus+{\bf w})\cdot
({\bf v}_{\rm sat}+{\bf v}_\oplus+{\bf w})\ .
\label{eq:8}
\end{equation}
This involves a term proportional to ${\bf v}_\oplus\cdot{\bf v}_{\rm
sat}$ which leads to observable effects, although
they are 12 times smaller than those proportional to
${\bf w}\cdot{\bf v}_{\rm sat}$, since
$v_\oplus/c\approx 9.94\times 10^{-5}$ whereas $w/c\approx
1.22\times 10^{-3}$ \cite{S91}. Section V.B is devoted to the
effects involving $\alpha_1 v_\oplus$ that are neglected
in the rest of the paper. In particular, we show there that the term
proportional to ${\bf w}\cdot{\bf v}_\oplus$ in Eq. (\ref{eq:8})
leads to sizable deviations of the angular position of the satellite,
which could be used to constrain the value of $\alpha_1$
independently of the tests proposed in sections III and IV. We
finally summarize our results and give our conclusions in section VI.

\section{Secular variations of the orbital elements}
The preferred-frame effects described by the Lagrangians 
(\ref{eq:3}) and (\ref{eq:4}) are $\sim\alpha_1 v^2/c^2$ smaller
than the leading Newtonian interaction, and lead, over one orbital
period, to very small deviations of the position of a satellite.
Taking into account the present limits (\ref{eq:6})-(\ref{eq:7}),
one expects $\alpha_1$-induced displacements $|\delta{\bf
x}|\sim\alpha_1 w\, v_{\rm sat} r_{\oplus{\rm sat}}/2c^2\alt 75\
\mu{\rm m}$ (for $r_{\oplus{\rm sat}}\sim 2 R_\oplus$), too small to
be observable with present or foreseeable techniques. Fortunately,
some of these perturbations build up beyond one orbital period, and
can thus be enhanced to an observable level if one waits for a
sufficiently large number of periods. We restrict our attention to
such effects in this paper, {\it i.e.} we concentrate on {\it
secular\/} variations of the orbital elements. Concentrating on
secular effects has also the advantage of freeing us from the
coordinate ambiguities present in orbital-period effects\cite{F1bis}.

When taking into account only the Newtonian potential generated by a
spherical Earth $Gm_\oplus m_{\rm sat}/r_{\oplus{\rm sat}}$, the
satellite's orbit is determined by six constants of motion: its
semi-major axis $a$, its eccentricity $e$, its inclination $I$ with
respect to the Earth's equatorial plane, the angle $\Omega$ between 
a direction of reference $(\alpha,\delta)=(0,0)$ \cite{F2} and the
ascending node, the angle $\omega$ between this ascending node and the
perigee, and finally the quantity $\sigma$ entering the mean anomaly
$\ell\equiv \int_0^t n(t') dt'+\sigma$, where
$n\equiv 2\pi/P = (Gm_\oplus/a^3)^{1/2}$ is the orbital frequency.
[Numerically $P = (a/R_\oplus)^{3/2}\times 1.406\ {\rm h}$, when
using $R_\oplus = 6.371\times 10^6\ {\rm m}$, $Gm_\oplus =
3.986\times 10^{14}\ {\rm m}^3{\rm s}^{-2}$.]
To help visualizing the meaning of the angle $\sigma$, it is useful
to note that, in the limiting case of a small eccentricity $e$, the
mean anomaly $\ell$ can be identified with the angular position of the
satellite, so that $\sigma$ can be viewed as the angle between the
perigee and the satellite at $t=0$. Figure 1 summarizes this
notation. It will also be useful in the following to define an
orthonormal triad $({\bf a}, {\bf b},{\bf c})$, where $\bf a$ is
directed towards the ascending node and ${\bf c}={\bf a}\times{\bf
b}$ is in the direction of the orbital angular momentum, {\it i.e.}
orthogonal to the orbital plane. (Note that this triad is {\it not\/}
the one used in references \cite{DEF2,DS91}, where $\bf a$ was
directed towards the periastron.)

We find it convenient to use the method of variation of the elements,
as described for instance in Ref. \cite{BC}, to derive the secular
variations of $a$, $e$, $I$, $\sigma$, $\omega$ and $\Omega$. For
more generality, let us consider a perturbed two-body system
$(m_A,m_B)$, define $M\equiv m_A+m_B$, $X_A\equiv m_A/M$, $X_B\equiv
m_B/M$, and write the Lagrangian as $L=L_0+MX_AX_B R$ where $L_0$
contains, besides the kinetic terms, only the Newtonian potential
between spherical bodies $Gm_Am_B/r_{AB}$, and
$R$ is the disturbing function containing all corrective terms due
to asphericities, tidal forces, relativistic effects, \dots
We shall not consider here the motion of the center of mass of the
two-body system \cite{F3}, but concentrate upon the equations
satisfied by the elements of the {\it relative\/} orbit ${\bf
x}_A-{\bf x}_B$. These derive directly from $R$, and read \cite{BC}
\begin{mathletters}
\label{eq:9}
\begin{eqnarray}
{da\over dt} = && {2\over na}\ {\partial R\over\partial\sigma}\ ,
\label{eq:9a}\\
{de\over dt} = && {1-e^2\over na^2e}\ {\partial R\over\partial\sigma}
-{(1-e^2)^{1/2}\over na^2e}\ {\partial R\over\partial\omega}\ ,
\label{eq:9b}\\
{dI\over dt} = && {\cot I\over na^2(1-e^2)^{1/2}}\ {\partial
R\over\partial\omega} -{1\over na^2(1-e^2)^{1/2}\sin I}\ {\partial
R\over\partial\Omega}\ , \label{eq:9c}\\
{d\sigma\over dt} = && -{2\over na}\left[{\partial R\over\partial
a}\right] -{1-e^2\over na^2e}\ {\partial R\over\partial e}\ ,
\label{eq:9d}\\
{d\omega\over dt} = && {(1-e^2)^{1/2}\over
na^2e}\ {\partial R\over\partial e} -{\cot I\over
na^2(1-e^2)^{1/2}}\ {\partial R\over\partial I}\ , \label{eq:9e}\\
{d\Omega\over dt} = && {1\over na^2(1-e^2)^{1/2}\sin I}\ {\partial
R\over\partial I}\ \cdot \label{eq:9f}
\end{eqnarray}
\end{mathletters}
The brackets in the right hand side of Eq.~(\ref{eq:9d}) indicate
that the $a$-differentiation is effected keeping the mean anomaly
fixed ({\it i.e.} ignoring the implicit $a$-dependence contained in
$\int n\, dt$). When working to first order in $R$, the
orbital elements can be replaced by their constant zeroth order values
on the right-hand-sides. Then the secular
variations of the orbital elements are simply obtained by replacing
in equations (\ref{eq:9}) $R$ by its average value
$\langle R\rangle$, computed over one unperturbed Newtonian orbit
[More precisely, one takes the average over the angle $\ell$,
keeping the other elements fixed]. It is straightforward to integrate
the terms proportional to
$1/c^2$ in the two-body versions of the Lagrangians (\ref{eq:2}) and
(\ref{eq:3}). Decomposing the `absolute' velocities as
${\bf v}^0_{A,B}={\bf v}_{A,B}+{\bf w}_{\rm CM}$, where ${\bf
w}_{\rm CM}$ is the velocity of the center-of-mass of the system
$(A,B)$ with respect to the preferred rest frame, we find
the contribution
\begin{eqnarray}
\langle R_{\alpha_1,\beta,\gamma}\rangle = &&-{\alpha_1\over
2}\left({GM\over ac^2}\right)c^2\left({w_{\rm CM}\over
c}\right)^2\nonumber\\
&&+{\alpha_1\over 2}\left({GM\over
ac^2}\right)^{3/2}c^2\,(X_A-X_B)\,{[{\bf w}_{\rm CM}/c,{\bf c},{\bf
e}]\over 1+(1-e^2)^{1/2}}\nonumber\\
&&+\left({GM\over
ac^2}\right)^2c^2\left({2\gamma-\beta+2+\alpha_1X_A X_B\over
(1-e^2)^{1/2}}-{8\gamma+7-X_A X_B(1-4\alpha_1)\over
8}\right),\label{eq:10}
\end{eqnarray}
where the square brackets denote the mixed product $({\bf
w}_{\rm CM}/c)\cdot({\bf c}\times{\bf e})$, the vector $\bf e$ being
the Lagrange-Laplace (-Runge-Lenz) eccentricity vector with norm $e$,
directed towards the periastron of body $A$. The first line comes
{}from the term proportional to $w_{\rm CM}^2$ in $L_{\alpha_1}$
(\ref{eq:3}), and the second from the term involving ${\bf w}_{\rm
CM}\cdot({\bf v}_A+{\bf v}_B)$. The third line is responsible for the
usual relativistic perigee advance and comes from the
$1/c^2$ terms of the Lagrangian (\ref{eq:1}), with extra
contributions due to the ${\bf v}_A\cdot{\bf v}_B$ term of
$L_{\alpha_1}$. We have not included in Eq. (\ref{eq:10}) the
contribution of $\alpha_2$, relying on the limit (\ref{eq:5}) to
consider that this type of preferred effects is already plausibly
excluded.

Up to here, we have considered a two-body system with arbitrary
masses $(m_A,m_B)$ for generality's sake. In the particular case of
artificial Earth satellites, one can however neglect $m_{\rm sat}$
with respect to $m_\oplus$, and therefore write $M\approx m_\oplus$,
$X_A = X_{\rm sat}\approx 0$, $X_B = X_\oplus\approx 1$. Hence, we
see from the last line in Eq. (\ref{eq:10}) that the additional
contribution to the relativistic perigee advance generated by
$\alpha_1$ is completely negligible. As for the first term on the
right hand side of Eq. (\ref{eq:10}), it only contributes to the
evolution of the element $\sigma$ and will be studied in section V.B.
In the next two sections we concentrate on the evolution of the other
elements, generated by the second term on the r.h.s. of Eq.
(\ref{eq:10}).

The contribution (\ref{eq:10}) should be added to the usual
Newtonian multipolar and tidal perturbations, notably the effect of
the Earth's quadrupolar moment ($J_2$) and the tidal forces
due to the Moon (denoted as $\Bbb C$) and the Sun (denoted as
$\odot$), which are the dominant ones. For analytical simplicity,
we shall not consider any other Newtonian perturbation in the present
paper. Our estimates of the measurability of $\alpha_1$ will
correspondingly be only indicative. To get reliable estimates of the
bounds on $\alpha_1$ which could be experimentally obtained, it
would be necessary to perform detailed numerical
simulations taking into account all known Newtonian and general
relativistic effects, as well as all modelizable sources of noise.
[There exist in particular resonances between  several multipolar
moments of the Earth which have significant effects on long times.]

The secular contribution due to the Earth's oblateness can be written
down very easily
\begin{equation}
\langle R_{J_2}\rangle = J_2\ {Gm_\oplus R_\oplus^2\over a^3}
\ {2-3\sin^2I\over4(1-e^2)^{3/2}}
\ \cdot
\label{eq:11}
\end{equation}
The contribution of tidal forces is more complicated (even if one
neglects the eccentricities of the Earth's and the Moon's orbits, as
well as their inclinations with respect to the Earth's equatorial
plane) because it involves explicitly the angular positions of the
Moon and the Sun. However, it can be simplified very much 
if the variations of the orbital elements (\ref{eq:9}) are studied on
longer time scales, more precisely if one averages over the orbital
periods of the Moon around the Earth ($\sim$ one month) and of the
Earth around the Sun (one year). The variations of the satellite's
orbital elements after such averagings can then be derived from
\begin{equation}
\langle\langle R_{\rm Tides}\rangle\rangle = 
N^2\,{a^2\over8}\,\left[ 2+3e^2-3\sin^2I\,(1-e^2+5e^2\sin^2\omega)
\right]\ ,
\label{eq:14}
\end{equation}
where
\begin{equation} N^2\equiv {Gm_{\Bbb C}\over r_{\oplus{\Bbb C}}^3}+
{Gm_\odot\over r_{\oplus\odot}^3}\ \cdot
\label{eq:13}
\end{equation}
[The notation $N$ is reminiscent of the fact that
$n_\oplus\equiv(Gm_\odot/r_{\oplus\odot}^3)^{1/2}$ is the orbital
frequency of the Earth around the Sun; note however that
$(Gm_{\Bbb C}/r_{\oplus{\Bbb C}}^3)^{1/2}$ is {\it not\/} the orbital
frequency of the Moon around the Earth, which would involve
$m_\oplus+m_{\Bbb C}$ instead of $m_{\Bbb C}$.]
Beware that the above expression (\ref{eq:14}) should not be used to
derive variations with periods $\alt$ one year.

The secular variations of the satellite's orbital elements can now
be easily derived from Eqs. (\ref{eq:9}). They tell us that the
semi-major axis undergoes no secular variation ($\langle \dot
a\rangle=0$), and that the change of the inclination is negligible in
the limit of a small eccentricity [$\langle dI/dt\rangle=O(\alpha_1
e)+O(e^2)$]. We shall study the equations satisfied by $e$, $\omega$
and $\Omega$ in sections III and IV, and see that they can be
rewritten more compactly as a vectorial equation for the eccentricity
vector $\bf e$ in the limit $e\ll 1$. Section V.B will be devoted to
the equation satisfied by $\sigma$.

Finally, it should be noted that in the main body of the paper we
always deal with the motion of a satellite as seen in a (locally
inertial) geocentric frame. The $\alpha_1$-dependent effects linked
to the connection between such a geocentric frame and a global,
barycentric one will be briefly discussed in the concluding section.

\section{Competition between tidal forces and Earth's oblateness for
equatorial orbits}
The aim of this section is to show on the simple example of
equatorial orbits how small divisors can enhance the preferred-frame
effects associated to $\alpha_1$. [As said above, we neglect the
eccentricities of the Earth's and the Moon's orbits, as well as
their inclinations with respect to the Earth's equatorial plane.] The
secular variations of the satellite's orbital elements are then given
by equations (\ref{eq:9}), where
$\langle R\rangle$ is the sum of Eqs. (\ref{eq:10}), (\ref{eq:11}) and
(\ref{eq:14}) with $I=0$. However, the angular position of the
ascending node is no longer well-defined as
$I\rightarrow 0$, and equations (\ref{eq:9e}) and (\ref{eq:9f}) become
formally singular. As is well known \cite{BC}, this singularity is
fictitious and taken care of by considering the evolution of the
angle $\widetilde\omega\equiv\omega+\Omega$ between the direction of
reference $(\alpha,\delta)=(0,0)$ and the perigee, which stays
well-defined in the limit $I\rightarrow 0$. One finds
\begin{eqnarray}
\left\langle{d\widetilde\omega\over dt}\right\rangle
= &&{(1-e^2)^{1/2}\over n a^2 e}\ {\partial \langle
R\rangle\over\partial e}\nonumber\\
= &&\dot{\widetilde\omega}_N -{\alpha_1\over 4}\ {n^2 a\over c}\ 
{[{\bf w}_\oplus/c, {\bf c}, {\bf e}]\over e^2} +O(\alpha_1 e)
\ , \label{eq:15}
\end{eqnarray}
where
\begin{equation}
\dot{\widetilde\omega}_N\equiv
n\left[{3\over 2}\ J_2 \left({R_\oplus\over a}\right)^2
+{3\over 4}\ {N^2\over n^2} +O(e^2)\right]
\label{eq:16}
\end{equation}
is the Newtonian perigee advance due to the Earth's quadrupolar
moment and tidal forces, and ${\bf w}_\oplus\equiv {\bf w}+{\bf
v}_\oplus$ the velocity of the Earth with respect to the preferred
frame, ${\bf v}_\oplus$ denoting as in section I the orbital
velocity of the Earth around the Sun. [It is useful to quote here the
numerical value of ${3\over 2}\,n\,J_2\,(R_\oplus/a)^2 =
(R_\oplus/a)^{7/2}\times 2.02\times 10^{-6}\ {\rm s}^{-1}\approx
(R_\oplus/a)^{7/2}\times 2\pi/(0.10\ {\rm yr})$, which will appear
again in the following sections as the characteristic frequency for
the Newtonian variations of the orbital elements $\omega$ and
$\Omega$; see notably Eqs. (\ref{eq:26}) and (\ref{eq:28}).] Equation
(\ref{eq:15}) together with the one satisfied by $e$
\begin{equation}
\left\langle{de\over dt}\right\rangle
= -{\alpha_1\over 4}\ {n^2 a\over c}\left({{\bf w}_\oplus\over
c}\cdot{{\bf e}\over e}\right) +O(\alpha_1 e^2)
\label{eq:17}
\end{equation}
can be rewritten as a simple vectorial equation for the
eccentricity vector $\bf e$ in the limit $e\ll 1$
\begin{equation}
\left\langle{d{\bf e}\over dt}\right\rangle
={\bf c}\times\left[
\dot{\widetilde\omega}_N\ {\bf e}+({\bf
k}+\bbox{\kappa}_\oplus)\times{\bf c}\right] +O(\alpha_1 e^2)+O(e^2)
\ ,
\label{eq:18}
\end{equation}
where
\begin{equation}
{\bf k} \equiv -\alpha_1\ {Gm_\oplus\over 4a^2c^2}\
{\bf w}
\label{eq:19}
\end{equation}
is a constant vector (beware that Ref. \cite{DEF2} defines a vector
$\bf k$ which is equal to twice this value) and
\begin{equation}
\bbox{\kappa}_\oplus(t)
\equiv -\alpha_1\ {Gm_\oplus\over 4a^2c^2}\ {\bf v}_\oplus
\label{eq:19bis}
\end{equation}
a yearly varying one. In the present and following sections, we
concentrate on the effects of the constant vector $\bf k$, leaving
to section V.A a study of the effects associated to
$\bbox{\kappa}_\oplus(t)$. Since we are considering equatorial orbits
in this section, the vector $\bf c$ orthogonal to the orbital plane
is the constant unitary vector parallel to the Earth's polar axis,
and therefore ${\bf c}\times({\bf k}\times{\bf c})$ is a {\it
constant\/} vector, namely the projection of $\bf k$ onto the
equatorial plane. Hence a
$\alpha_1$-type violation of local boost invariance has the
consequence of adding a constant forcing term in the time evolution
of the eccentricity vector which, if it were alone, would secularly
`polarize' the orbit in
the direction of the equatorial projection of $\bf w$. However, the
Newtonian precession term $\dot{\widetilde\omega}_N$ cuts off the
build up of this polarizing term and deflects it by $90^\circ$ in a
gyroscope-like way. More precisely, the solution of Eq. (\ref{eq:18})
can be written as the vectorial superposition
\begin{equation}
{\bf e} = {\bf e}_N(t) +{\bf e}_F
\ ,
\label{eq:20}
\end{equation}
where ${\bf e}_N(t)$ is a vector of constant norm rotating with
angular frequency $\dot{\widetilde\omega}_N$ in the orbital plane
(usual Newtonian perigee advance), and ${\bf e}_F$ is the fixed
polarizing contribution due to the preferred-frame effects we are
studying, namely
\begin{equation}
{\bf e}_F\equiv
{{\bf c}\times{\bf k}\over\dot{\widetilde\omega}_N}
\ \cdot
\label{eq:21}
\end{equation}
Note the factor $(\dot{\widetilde\omega}_N)^{-1}$ in Eq. (\ref{eq:21})
which is the first appearance of a small divisor. The solution
(\ref{eq:20}) is formally identical to the one found for binary
pulsars in Ref. \cite{DEF2}, the only difference being that
$\dot{\widetilde\omega}_N$ replaces the relativistic periastron
advance. The same kind of polarizing terms in the time evolution of
$\bf e$ has also been pointed out previously in
\cite{DS91}, in the totally different context of equivalence
principle violation in binary systems.

In geometrical terms, Eq. (\ref{eq:20}) means that the eccentricity
vector ${\bf e}(t)$ traces out, during its time evolution (after
averaging over an orbital period), a circle centered around ${\bf
e}_F$. If a large enough segment of this circle (say about a
quarter) can be monitored during the experiment, it should be
possible to measure the position of its center, {\it i.e.} ${\bf
e}_F$, with about the same precision that individual measurements of
the eccentricity vector \cite{F4}. In turn, the precision of the
measurement of $\bf e$ is related to the precision $\sigma_{\bf x}$
with which one can measure the satellite's position: roughly one
expects $\sigma_{\bf e}\approx \sigma_{\bf x}/a$. It is convenient
to work with quantities homogeneous to displacements. Therefore we
shall measure the $\alpha_1$-perturbations of $\bf e$ in terms of
\begin{equation}
\delta\rho = a e_F=
{\alpha_1\over 3 c^2}\ {|{\bf c}\times{\bf w}|\ (Gm_\oplus)^{3/2}
\over
2 J_2 G m_\oplus R_\oplus^2 a^{-5/2} + N^2 a^{5/2}
}
\ ,
\label{eq:22}
\end{equation}
which represents (when $e_F<e_N$, as expected from
$e_F\leq \alpha_1 \times 5.7 \times 10^{-5} < 3 \times 10^{-8}$)
the amplitude of the secular change in the distance to the perigee
$\rho = a(1-|{\bf e}|)$. From the arguments just given, equating
$\delta \rho$ with the position measurement precision $\sigma_{\bf
x}$ should yield an estimate of the precision with which $\alpha_1$
can be measured over a time span $T/4$, where
\begin{equation}
T \equiv {2\pi\over\dot{\widetilde\omega}_N}
\ \cdot
\label{eq:22bis}
\end{equation}
The denominator of the expression (\ref{eq:22}) is a function of the
semi-major axis $a$, and is minimized for the value
\begin{eqnarray}
a_{\rm optimal}= &&(2J_2Gm_\oplus R_\oplus^2/N^2)^{1/5}
\nonumber\\
\approx &&7.66 R_\oplus \approx 4.88\times 10^7 {\rm m}
\label{eq:23}
\end{eqnarray}
(which corresponds to an orbital period $\sim 30$ hours). In Eq. 
(\ref{eq:23}), we have used the numerical data $J_2=1.08263\times
10^{-3}$,
$Gm_\oplus/R_\oplus c^2 = 6.97\times 10^{-10}$,
$R_\oplus=6.371\times 10^6\ {\rm m}$, and $N=3.56\times10^{-7}\ {\rm
s}^{-1}$. The maximal value of $\delta \rho$ (for equatorial
orbits) is thus obtained for this value of the semi-major axis, and
reads
\begin{eqnarray}
\delta\rho_{\rm max}= &&\alpha_1{Gm_\oplus\over R_\oplus c^2}\
{c\over 6N(2J_2)^{1/2}}\ {w_{\rm equatorial}\over c}
\nonumber\\
\approx &&\alpha_1\times 2.54\times 10^5\ {\rm cm}
\ , \label{eq:24}
\end{eqnarray}
where the norm ${w/c}= 1.22\times 10^{-3}$ and the direction
$(\alpha,\delta)=(11.2\ {\rm h}, -7^\circ)$ of $\bf w$ have been
extracted from the results of COBE \cite{S91}.

Present technologies make it possible to measure the position of a
satellite down to $\sigma_{\bf x}\sim 1\ {\rm cm}$, by using either
laser ranging (as for the Laser Geodynamical Satellites `LAGEOS') or
Global Positioning System (GPS) receivers onboard. The displacement
(\ref{eq:24}) could therefore lead in principle to limits on
$\alpha_1$ of the order of
$4\times 10^{-6}$, {\it i.e.} two orders of magnitude tighter than
the present bounds (\ref{eq:6}) and (\ref{eq:7}). It is to be noted
that a geostationary satellite ($a\approx 6.62\,R_\oplus$,
${\rm period} \approx 23\ {\rm h}\ 56\ {\rm min}$) is near optimal.
It yields $\delta\rho\approx \alpha_1\times 2.4\times 10^5\ {\rm
cm}$, leading us to expect a precision $\alpha_1\sim 4.2\times
10^{-6}$. Clearly, numerical simulations taking into account all
known Newtonian perturbations and modelizable non-conservative forces
should be performed in order to get more realistic estimates of the
precision which can be reached.

This simple case of equatorial orbits already exhibits one of the
characteristic behaviors of preferred-frame effects enhanced by
small divisors: the duration of
the experiment must be large enough to take advantage of the small
divisor effect. For instance, the time span of the experiment should
be (at least) $T/4 = \pi/2\dot{\widetilde\omega}_N\sim 15.4\ {\rm
yr}$ in the case of the optimal orbit (\ref{eq:23}), and $\sim 12.4\
{\rm yr}$ for a geostationary satellite.

Let us conclude this section by a comparison of the result
(\ref{eq:24}) with the one corresponding to the best
drag-free satellite we know: the Moon itself. Still neglecting the
inclination of its orbit and that of the ecliptic with respect to the
Earth's equatorial plane, we can repeat all of the above discussion,
except that in Eq. (\ref{eq:13}) only the tidal forces due to the Sun
should be taken into account. Actually, higher-order terms in
these tidal forces are not small in the particular case of the Moon
(because its orbital period is non-negligible as compared to one
year), and the coefficient $N^2$ of Eq. (\ref{eq:14}) should be
replaced by $2.043\ n_\oplus^2$ instead of merely $n_\oplus^2\equiv
Gm_\odot/r_{\oplus\odot}^3$ to yield the correct magnitude
of $\dot{\widetilde\omega}_N$ \cite{BC}. The amplitude of the
secular oscillation of the Moon's perigee distance is then given by
the same formula (\ref{eq:22}) as above with
$a=r_{\oplus{\Bbb C}}=60.3\ R_\oplus$. The Earth's oblateness gives
then a negligible contribution to the denominator of
$\delta\rho$ in Eq. (\ref{eq:22}), and one finds $\delta\rho_{\Bbb
C} \approx \alpha_1\times 4.5\times 10^3\ {\rm cm}$, with period
$T\approx 8.72\ {\rm yr}$. This effect was
first discussed (in a different guise) in Ref. \cite{N}. Since Lunar
Laser Ranging (LLR) allows one to determine the Earth--Moon distance
within 2 or 3 centimeters, we expect that an analysis of LLR data
could at best limit $\alpha_1$ at the $\sim 5\times 10^{-4}$ level,
{\it i.e.} not better than the existing bounds
(\ref{eq:6}),(\ref{eq:7}). However, as this would represent a new,
independent test of preferred-frame effects, it would be interesting
to perform explicitly a multi-parameter fit of LLR data including the
contributions of
$\alpha_1$ (and for completeness $\alpha_2$, $\beta$, $\gamma$) to
the Lunar equations of motion.

\section{Non-equatorial orbits and arbitrarily small divisors}
In this section, we shall see that non-equatorial orbits (of any
altitude) can give rise to arbitrarily small divisors if the
inclination is near some special values. For reasons that will
appear clear below, we concentrate upon rather low satellites, for
which tidal forces are negligible compared to the influence of the
Earth's quadrupolar moment. For this reason we do not take
into account the disturbing function (\ref{eq:14}) in the present
section, in order to avoid unnecessarily technical calculations.
However, we show at the end of this section how the equations
satisfied by the orbital elements can be solved when tidal forces are
not neglected.

The main difference between equatorial and non-equatorial satellites
is that the orbital plane is no longer fixed when $I\neq 0$.
Indeed, equation (\ref{eq:9f}) shows that it is precessing with a
constant angular frequency
\begin{equation}
\langle\dot\Omega\rangle = -{3\over 2}\ n\,
J_2\left({R_\oplus\over a}\right)^2 \cos I +O(\alpha_1
e)+O(e^2)+O(N^2) \ ,
\label{eq:26}
\end{equation}
which is not modified by a $\alpha_1$-type violation of local
Lorentz invariance in the limiting case of a small eccentricity. We
shall drop the angular brackets in the following, and denote
this constant precessing velocity simply by $\dot\Omega$. Like in
the previous section, the equations (\ref{eq:9b}) and (\ref{eq:9e})
satisfied by $e$ and $\omega$ can be rewritten as a simple vectorial
equation for the evolution of the components of $\bf e$ with respect
to the vectors $({\bf a},{\bf b})$, which are part of a frame
[defined at the beginning of section II and in Fig.~1] which rotates
around the Earth's polar axis with the angular velocity
(\ref{eq:26}):
\begin{equation}
\langle d'{\bf e}/dt\rangle=
{\bf c}\times\bigl[\dot\omega_N{\bf e}+({\bf
k}+\bbox{\kappa}_\oplus)\times{\bf c}\bigr] +O(\alpha_1
e^2)+O(e^2)+O(eN^2)
\ ,
\label{eq:27}
\end{equation}
where the prime in $d'/dt$ denotes a time derivative in the rotating
frame $({\bf a},{\bf b},{\bf c})$ and where
\begin{equation}
\dot\omega_N\equiv {3\over 4}\ n\, J_2\left({R_\oplus\over a}\right)^2
(4-5\sin^2 I)+O(e^2)+O(N^2)
\label{eq:28}
\end{equation}
is the Newtonian perigee advance due to the Earth's quadrupolar
moment, ${\bf k}$ and $\bbox{\kappa}_\oplus$ being the vectors
defined in Eqs. (\ref{eq:19}),(\ref{eq:19bis}). As in section III,
we shall neglect in this section $\bbox{\kappa}_\oplus$ with respect
to $\bf k$. Since the projection
${\bf c}\times({\bf k}\times{\bf c})$ of $\bf k$ onto the orbital
plane is no longer a constant vector because of the precession
$\dot\Omega$, the solution (\ref{eq:20}) of the previous section is
not valid for non-equatorial orbits. However, it is easy to solve
equation (\ref{eq:27}) which is just an inhomogeneous {\it linear\/}
differential equation in $\bf e$ (in the limiting case of a small
eccentricity). Let us look for a solution of the type
${\bf e}={\bf e}_N+\sum_i{\bf e}_i$, where ${\bf e}_N$ is the usual
(constant-norm) Newtonian eccentricity vector rotating in the
orbital plane $({\bf a},{\bf b})$ with angular frequency
$\dot\omega_N$, and where the ${\bf e}_i$'s are some constant-norm
vectors rotating in the orbital plane with constant angular
frequencies
$\dot\omega_i$ to be determined. The time derivative of $\bf e$
then reads
\begin{eqnarray}
d'{\bf e}/dt= &&{\bf c}\times\left(\dot\omega_N\ {\bf
e}_N+\sum_i\dot\omega_i\ {\bf e}_i\right)\nonumber\\
= &&{\bf c}\times\left(\dot\omega_N\ {\bf
e}+\sum_i\,(\dot\omega_i-\dot\omega_N)\,{\bf e}_i\right)
\ .
\label{eq:29}
\end{eqnarray}
This has precisely the form of equation (\ref{eq:27}), with ${\bf
k}\times{\bf c}$ having been decomposed as a sum of constant-norm
vectors $(\dot\omega_i-\dot\omega_N){\bf e}_i$ rotating with
constant angular frequencies $\dot\omega_i$ in the orbital plane. 
Since ${\bf k}\times{\bf c}$ has a periodic motion in the orbital
plane (with period $2\pi/\dot\Omega$), it obviously admits such a
decomposition:
\begin{equation}
{\bf k}\times{\bf c} = -\alpha_1\ {Gm_\oplus\over4 a^2 c^2}\ w\,
({\bf K}_+ +{\bf K}_0 + {\bf K}_-) \ ,
\label{eq:30}
\end{equation}
where
\begin{mathletters}
\label{eq:31}
\begin{eqnarray}
{\bf K}_\pm\equiv &&\cos\delta\ {1\pm\cos I\over 2}\ \bigl[\mp
\sin(\Omega-\alpha)\ {\bf a}-\cos(\Omega-\alpha)\ {\bf b}\bigr]
\ ,
\label{eq:31a} \\
{\bf K}_0\equiv &&\sin\delta\, \sin I\ {\bf a}
\ ,
\label{eq:31b}
\end{eqnarray}
\end{mathletters}
$(\alpha, \delta)$ denoting the right-ascension and declination of
$\bf w$. ${\bf K}_\pm$ are rotating with angular frequencies
$\mp\dot\Omega$ in the orbital plane, whereas ${\bf K}_0$ is a
constant vector in this plane (directed towards the ascending node).
The solution of Eq. (\ref{eq:27}) can therefore be written simply as
\begin{eqnarray}
{\bf e}= &&{\bf e}_N+{\bf e}_+ +{\bf e}_0 +{\bf e}_- \nonumber\\
= &&{\bf e}_N+\alpha_1\ {Gm_\oplus\over 4 a^2 c^2}\ w \left(
{{\bf K}_+\over\dot\omega_N+\dot\Omega} +
{{\bf K}_0\over\dot\omega_N} +
{{\bf K}_-\over\dot\omega_N-\dot\Omega}
\right)
\ .
\label{eq:32}
\end{eqnarray}
This generalizes the solution (\ref{eq:20}) obtained above for
equatorial orbits. [In the limit $I\rightarrow 0$, the term
involving ${\bf K}_+$ becomes equal to the vector ${\bf e}_F$ of Eq.
(\ref{eq:21}), in which
$\dot\omega_N+\dot\Omega$ was denoted $\dot{\widetilde\omega}_N$.]
In geometrical terms, Eq. (\ref{eq:32}) means that the eccentricity
vector ${\bf e}(t)$ undergoes a kind of epicyclic motion in the
$({\bf a},{\bf b})$ plane: it moves (with angular velocity
$\dot\omega_N$) along a circle of radius $|{\bf e}_N|$ whose center
moves itself (with a non-uniform angular velocity determined by
$\dot\Omega$) on an ellipse $\bigl({\bf e}_+(t)+{\bf e}_-(t)\bigr)$
centered around ${\bf e}_0$. Hence we can now distinguish three
different contributions to the $\alpha_1$-induced secular
oscillation of the distance to the perigee:
\begin{mathletters}
\label{eq:33}
\begin{eqnarray}
\delta\rho_\pm= && a e_\pm =
\alpha_1\ {Gm_\oplus\over 4 a c^2}\ w\cos\delta\ {(1\pm\cos
I)/2\over\dot\omega_N\pm\dot\Omega}
\nonumber\\
= &&A\cos\delta\ {(1\pm\cos I)/2\over4-5\sin^2 I\mp2\cos I}
\ , \label{eq:33a}\\
\delta\rho_0= &&a e_0 =
\alpha_1\ {Gm_\oplus\over 4 a c^2}\ w\sin\delta\ {\sin
I\over\dot\omega_N}\nonumber\\
= &&A\sin\delta\ {\sin I\over 4-5\sin^2 I}
\ , \label{eq:33b}
\end{eqnarray}
\end{mathletters}
where
\begin{eqnarray}
A \equiv &&\alpha_1\left({a\over
R_\oplus}\right)^{5/2}\left({Gm_\oplus\over R_\oplus
c^2}\right)^{1/2}{R_\oplus\over 3 J_2}\ {w\over c}
\nonumber\\
\approx &&\alpha_1\, ({a/ R_\oplus})^{5/2}\times 6316\
{\rm cm.}
\label{eq:34}
\end{eqnarray}
Figure 2 displays the amplitude of these displacements as functions
of the inclination. They clearly exhibit poles (or `resonances') for
six particular values of
$I$: $\delta\rho_+$ diverges for $\dot\omega_N+\dot\Omega=0$, {\it
i.e.} for $I=46.38^\circ$ or $106.85^\circ$; $\delta\rho_0$ diverges
for $\dot\omega_N=0$, {\it i.e.} for
$I=\sin^{-1}(2/\sqrt{5})=63.43^\circ$ or $116.57^\circ$;
and $\delta\rho_-$ diverges for $\dot\omega_N-\dot\Omega=0$, {\it
i.e.} for $I=73.15^\circ$ or $133.62^\circ$. (Note that the special
value $I_0\equiv\sin^{-1}(2/\sqrt{5})$ coincides with the well-known
Newtonian critical inclination for $J_2$-effects \cite{BC}.) As in
the equatorial case discussed above, the price to pay for taking
advantage of the small divisors arising near these poles is the need
for a long observation period, inversely proportional to the
corresponding small divisor. Indeed, the dephasing periods between
the Newtonian eccentricity vector ${\bf e}_N$ and the contributions
${\bf e}_\pm$,${\bf e}_0$ proportional to $\alpha_1$ read
\begin{mathletters}
\label{eq:35}
\begin{eqnarray}
T_\pm= &&{2\pi\over \dot\omega_N\pm\dot\Omega} = {B\over
4-5\sin^2 I\mp 2\cos I}
\ ,
\label{eq:35a} \\
T_0 = &&{2\pi\over\dot\omega_N} = {B\over 4 - 5\sin^2 I}
\ , \label{eq:35b}
\end{eqnarray}
\end{mathletters}
where
\begin{eqnarray}
B \equiv
&&\left({a\over R_\oplus}\right)^{7/2} \left({Gm_\oplus\over
R_\oplus c^2}\right)^{-1/2}{8\pi\over 3 J_2}\ {R_\oplus\over c}
\nonumber\\
\approx &&({a/ R_\oplus})^{7/2}\times 0.1974\
{\rm yr.}
\label{eq:36}
\end{eqnarray}
[Near each resonance $\dot\omega_N\pm\dot\Omega=0$, one of the vectors
${\bf e}_\pm$ becomes infinitely large with respect to the other, and
the ${\bf e}_++{\bf e}_-$ ellipse degenerates to a circle described
with angular velocity $\mp\dot\Omega\approx\dot\omega_N$.]

In order to compare the merits of the different resonances, it is
useful to define as a figure of merit the ratio of the corresponding
perigee displacement $\delta\rho$ by the typical (minimal) time $T/4$
needed for the observation:
\begin{mathletters}
\label{eq:37}
\begin{eqnarray}
{4\delta\rho_\pm\over T_\pm} = &&4\ {A\over B}\ \cos\delta\
{1\pm\cos I\over 2} \nonumber\\
= &&\alpha_1\ {R_\oplus\over a}\times 1.073\times10^5\
{\rm cm.yr}^{-1}\quad{\rm for}\ I=46.38^\circ\ {\rm and}\ 133.62^\circ
\ ,
\label{eq:37a} \\
= &&\alpha_1\ {R_\oplus\over a}\times 4.509\times10^4\
{\rm cm.yr}^{-1}\quad{\rm for}\ I=73.15^\circ\ {\rm and}\ 106.85^\circ
\ , \label{eq:37b}\\
{4\delta\rho_0\over T_0} = &&4\ {A\over B}\ \sin\delta\, \sin I
\nonumber\\
= &&\alpha_1\ {R_\oplus\over a}\times 1.395\times10^4\
{\rm cm.yr}^{-1}\quad{\rm for}\ I=63.43^\circ\ {\rm and}\ 116.57^\circ
\ . \label{eq:37c}
\end{eqnarray}
\end{mathletters}
These results underline that the inclinations around $46.38^\circ$
or $133.62^\circ$ have the capability to give the largest effects in a
given observational time, as confirmed by the width of the
resonances in Fig. 2. Equations (\ref{eq:37}) also show that
low orbits (say $R_\oplus<a\alt 2R_\oplus$) seem preferable in that
they give larger $\delta\rho$'s in a given observational time. [It is
easy to see that the $a^{-1}$ dependence of $\delta\rho/T$ still
applies for high, tidally perturbed orbits. This justifies our
concentrating on low orbits, with negligible tidal effects.] However,
equations (\ref{eq:35}),(\ref{eq:36}) show that the inclination $I$
must be fixed with a high precision, typically within less than one
arc-minute if one wishes to make the fullest use of the
observational time $T/4$ which, for practical reasons, will probably
not exceed $\sim 10$ years. The needed inclination can be
computed thanks to the following asymptotic formulae:
\begin{mathletters}
\label{eq:38}
\begin{eqnarray}
| I-I_{\rm pole} |\approx &&
\left({a\over R_\oplus}\right)^{7/2} 3.19^\circ\ {1\ {\rm yr}\over
T_\pm}\quad{\rm for}\ I_{\rm pole}=46.38^\circ\ {\rm or}\
133.62^\circ \ ,
\label{eq:38a} \\
| I-I_{\rm pole} |\approx &&
\left({a\over R_\oplus}\right)^{7/2} 2.41^\circ\ {1\ {\rm yr}\over
T_\pm}\quad{\rm for}\ I_{\rm pole}=73.15^\circ\ {\rm or}\
106.85^\circ \ ,
\label{eq:38b} \\
| I-I_{\rm pole} |\approx &&
\left({a\over R_\oplus}\right)^{7/2} 2.83^\circ\ {1\ {\rm yr}\over
T_0}\quad{\rm for}\ I_{\rm pole}=63.43^\circ\ {\rm or}\
116.57^\circ \ .
\label{eq:38c}
\end{eqnarray}
\end{mathletters}
For instance a low satellite ($a\approx R_\oplus$) will give rise to
a perigee displacement
$\delta\rho_\pm\approx \alpha_1\times 10^6\ {\rm cm}$ in $T_\pm/4=10\
{\rm yr}$ if the inclination differs from the pole value
$43.38^\circ$ (or $133.62^\circ$) by only $3.19^\circ/40=4'47''$. An
error of one arc-minute on $I$ would change the observational time
and the perigee displacement by factors $\sim (0.8)^{\pm 1}$.

The six resonant values of the inclination are solutions of the
simple trigonometric equations $4-5\sin^2 I+2x\cos I=0$, where
$x\in\{-1,0,1\}$, and do not depend on any experimental data. They
will therefore enhance preferred-frame effects in the motion of the
natural satellites of the different planets in the solar system
(although tidal forces may not be negligible is some cases). However,
none of the known natural satellites' orbits has an inclination close
enough to one of the six poles. On the other hand, there are hundreds
of artificial Earth satellites, and many of them would allow one to
tighten the present bounds (\ref{eq:6}),(\ref{eq:7}) on $\alpha_1$ if
the evolution of their perigee could be tracked at the centimeter
level. For instance, the classes of satellites `GPS BII' and
`GOES' would typically allow one to measure perigee radial
displacements of
$\alpha_1\times3\times10^5\ {\rm cm}$ in an observational time of
$T/4\sim 15\ {\rm yr}$. Even if these satellites could be tracked
with sufficient precision, it would remain to see whether the effect
of non-gravitational forces would allow one to make full use of such
long data span. However, a statistical study of these $\sim
20$ satellites could probably allow a significant reduction of the
sources of errors. The same remarks can be formulated about the
classes of satellites `GPS' and `Glonass' whose perigees are deviated
by $\sim\alpha_1\times1.5\times10^5\ {\rm cm}$ in $T/4\sim 8\ {\rm
yr}$. [There are also many low ({\it i.e.} fast) satellites, like
`Starlette' or `Seasat', which give perigee oscillations of $\sim
\alpha_1\times3\times 10^4\ {\rm cm}$ in less than one year, but the
air drag is very large for such low satellites.] The first Laser
Geodynamical Satellite (LAGEOS I) is an interesting candidate, not
only because its position has been laser tracked for years at the few
centimeter level and because it is submitted to very small air drag,
but also because {\it several\/} contributions to the eccentricity
vector
$\bf e$ are large in its case: ${\bf e}_+$ gives a perigee deviation
of
$\alpha_1\times 4.2\times10^4\ {\rm cm}$ in $T_+/4\sim 1.9$ years,
${\bf e}_-$ a deviation of
$\alpha_1\times 2.0\times10^4\ {\rm cm}$ in $T_-/4\sim 5.3$ months,
and ${\bf e}_0$ a deviation of
$\alpha_1\times 8.8\times10^3\ {\rm cm}$ in $T_0/4\sim 1.2$ years.
The superposition of these three effects thus yields a complex
signal which should, hopefully, be distinguishable from other,
Newtonian, contributions. Unfortunately, the variations of the
eccentricity $e$ of this satellite are not very precisely
modelizable, as explained in Refs. \cite{SD80,MNF}: it undergoes
small oscillations of
$\pm 3\times10^{-7}$, which correspond to perigee deviations of
$\pm 4\ {\rm m}$. The recently launched second LAGEOS, which has
an even more favorable inclination, might provide another
interesting experimental probe of preferred-frame effects. Indeed,
a duration of only $T_+/4\sim 1.3\ {\rm yr}$ would suffice
to observe a radial displacement of its perigee by
$\delta\rho_+\sim \alpha_1\times 7\times 10^4\ {\rm cm}$. The best
tool for tightening the present bounds (\ref{eq:6}),(\ref{eq:7}) on 
$\alpha_1$ would be a (naturally or artificially) drag-free
satellite, launched on a favorably inclined orbit and tracked for
years at the centimeter level ({\it via\/} laser ranging or onboard
GPS receivers). Let us note that there are plans for launching in
the near future drag-free satellites with GPS receivers onboard:
Gravity Probe B (GPB) and the Satellite Test of the Equivalence
Principle (STEP).

For completeness, we now briefly discuss how the above solution
(\ref{eq:32}) for the eccentricity vector $\bf e$ is modified when
tidal forces (which are very small for low satellites) are taken into
account. The first (trivial) modification is of
course that the disturbing function (\ref{eq:14}) now gives a
contribution to the precession velocity
$\langle\dot\Omega\rangle$ (\ref{eq:9f}) of the orbital plane:
\begin{equation}
\langle\dot\Omega\rangle = -{3\over 4}\ n\, \cos I\, \bigl[2
J_2(R_\oplus/a)^2+N^2/n^2\bigr]+O(\alpha_1 e)+O(e^2)\ .
\label{eq:39}
\end{equation}
On the other hand, the effect of tidal forces in
equations (\ref{eq:9b}) and (\ref{eq:9e}) for $e$ and $\omega$ is
somewhat more involved [even after averaging over monthly and yearly
frequencies as explained in section II, Eq. (\ref{eq:14})]. The
vectorial equation satisfied by the eccentricity vector $\bf e$ in the
orbital plane reads now
\begin{equation}
\langle d'{\bf e}/dt\rangle = {\bf c}\times(\mu\ {\bf e}+{\bf
k}\times{\bf c})+\nu\ {\bf b}\times({\bf a}\times{\bf e})+O(\alpha_1
e^2)+O(e^3)
\ ,
\label{eq:40}
\end{equation}
where
\begin{equation}
\mu\equiv {3\over 4}\ n\, \bigl[J_2(R_\oplus/a)^2(4-5\sin^2
I)+2N^2/n^2\bigr]
\label{eq:41}
\end{equation}
is the analog of the angular frequency $\dot\omega_N$ of Eq.
(\ref{eq:27}), and
\begin{equation}
\nu\equiv{15\over 4n}\ N^2\sin^2 I
\label{eq:42}
\end{equation}
is an additional contribution due to tidal forces. 
The ratio $\nu/\mu$ can become positive and larger than unity
when the inclination $I$ is very close to the well-known
$J_2$-critical inclination $I_0\equiv \sin^{-1}(2/\sqrt{5})\approx
63.43^\circ$ or $116.57^\circ$ ({\it e.g.} $|I-I_0|\alt 4'$ for
$a\sim 2R_\oplus$, and $|I-I_0|\alt 8''$ for $a\sim R_\oplus$).
Equation (\ref{eq:40}) then formally exhibits an exponential blow up
of the Newtonian eccentricity vector. This indicates that our
simplified (linearized and time-averaged) treatment of the evolution
of $\bf e$ becomes inadequate. In the following, we restrict our
attention to the generic case $\nu/\mu\leq 1$. One finds that ${\bf
e}_N$ has no longer a constant norm, as opposed to Eq. (\ref{eq:20})
and (\ref{eq:32}) above, but that it is now moving on an ellipse,
whose axes are directed along $\bf a$ and $\bf b$ and have a ratio
$e_a/e_b=(1-\nu/\mu)^{1/2}$~:
\begin{equation}
{\bf e}_N = e_a\cos\left(\mu\sqrt{1-\nu/\mu}\ t+{\rm cst}\right)\ {\bf
a} +e_b\sin\left(\mu\sqrt{1-\nu/\mu}\ t+{\rm cst}\right)\ {\bf b}
\ .
\label{eq:43}
\end{equation}
[Note that this ellipse reduces to a circle in the particular case
of equatorial orbits ($I=0$) that we considered in section II.]
The solution of equation (\ref{eq:40}) can then be obtained in a
geometrical manner similar to the one used in Eqs.
(\ref{eq:29})--(\ref{eq:32}). Let ${\bf e}_i$ denote a
generalization of the Newtonian eccentricity vector ${\bf e}_N$ when
the frequency $\mu$ is replaced by an arbitrary constant $\mu_i$
such that $\nu/\mu_i\leq 1$, {\it i.e.} $\mu_i\leq 0$ or
$\mu_i\geq\nu$ (we shall see below that these values of $\mu_i$ are
sufficient for our purpose). Like ${\bf e}_N$ in Eq. (\ref{eq:43})
above, ${\bf e}_i$ is moving on an ellipse in
the orbital plane, with a non-uniform velocity which has the sign of
$\mu_i$, and such that its components along $\bf a$ and $\bf b$ are
oscillating with a constant frequency $\mu_i(1-\nu/\mu_i)^{1/2}$ and
have an amplitude ratio $e^i_a/e^i_b = (1-\nu/\mu_i)^{1/2}$. [The
particular cases of constant vectors along $\bf a$ or $\bf b$ are
obtained respectively for $\mu_i\rightarrow 0^-$ and $\mu_i = \nu$].
The time derivative of a linear combination ${\bf e}
\equiv {\bf e}_N+\sum_i{\bf e}_i$ then reads
\begin{equation}
d'{\bf e}/dt = {\bf c}\times\left(\mu\ {\bf
e}+\sum_i\,(\mu_i-\mu)\,{\bf e}_i\right)+\nu\ {\bf b}\times({\bf
a}\times{\bf e})
\ .
\label{eq:44}
\end{equation}
This has precisely the form of Eq. (\ref{eq:40}), with ${\bf
k}\times{\bf c}$ having been decomposed as a sum of vectors
$(\mu_i-\mu)\,{\bf e}_i$ moving on ellipses as described above. It is
easy to check that any vector having a periodic motion in the
orbital plane admits such a decomposition, and that the
corresponding $\mu_i$'s always satisfy the condition $\nu/\mu_i\leq
1$. In particular, one finds that the source term ${\bf k}\times{\bf
c}$ can be decomposed as in Eq. (\ref{eq:30}), where ${\bf K}_0$ is
still given by Eq. (\ref{eq:31b}), but where ${\bf K}_\pm$ have
more complicated expressions
\begin{equation}
{\bf K}_\pm\equiv 
\cos\delta\ {\dot\Omega/\mu_\mp + \cos I\over
2 - \nu/\mu_\mp}\left[
-\sin(\Omega-\alpha)\ {\bf
a}+{\mu_\pm\over\dot\Omega}\cos(\Omega-\alpha)\ {\bf b}\right]
\ ,
\label{eq:44bis}
\end{equation}
with
\begin{equation}
\mu_\pm\equiv \mp
\dot\Omega\sqrt{1+(\nu/2\dot\Omega)^2}+\nu/2
\ .
\label{eq:44ter}
\end{equation}
Note that these expressions for ${\bf K}_\pm$ reduce to those of
Eq. (\ref{eq:31a}) when $\nu=0$, {\it i.e.} when tidal forces are
neglected or for equatorial orbits. Hence the solution of Eq.
(\ref{eq:40}) can be written simply as
\begin{equation}
{\bf e}= {\bf e}_N+\alpha_1\ {Gm_\oplus\over 4 a^2 c^2}\ w
\left( {{\bf K}_+\over\mu-\mu_+} + {{\bf
K}_0\over\mu} + {{\bf K}_-\over\mu-\mu_-}
\right)
\ .
\label{eq:45}
\end{equation}
It is therefore of the same kind as Eq. (\ref{eq:32}),
and there still exist poles for some particular values of the
inclination, only slightly modified with respect to those of Fig. 2
for relatively low satellites. [For the same reason indicated above,
our treatment becomes inadequate when $\mu\rightarrow 0^+$, formally
corresponding to $(1-\nu/\mu)^{1/2}$ becoming large and pure
imaginary.]

\section{Perturbations due to the orbital velocity of the Earth
around the Sun}
\subsection{Perturbations of the eccentricity vector}
In the previous sections, we have neglected the orbital velocity
${\bf v}_\oplus$ of the Earth around the Sun, but we shall see below
that it can also lead to significant preferred-frame effects on
artificial satellites, although they will typically be 12 ($\approx
w/v_\oplus$) times smaller than those proportional to $w$. The
interest of effects involving ${\bf v}_\oplus$ is twofold. First of
all, we shall see that arbitrarily small divisors can enhance the
preferred-frame effects we are studying for {\it any\/} value of the
inclination $I$ of the satellite's orbit (if the semi-major axis $a$
is chosen appropriately) instead of the discrete resonances displayed
in Fig.~2. Moreover, the existence of a motion around the Sun (with
well-defined amplitude and phase) is known for sure to be part of
the `absolute' velocities entering preferred-frame effects. By
contrast, the identification of $\bf w$ with our velocity with
respect to the cosmic microwave background is a specific assumption.
Although this assumption is plausible on field-theoretical grounds,
the bounds derived on $\alpha_1$ in the literature
\cite{W81,NW72,WG76,H84,DEF2} and the discussions of the previous
sections of the present paper are strongly dependent on it. It
seems therefore of importance to determine limits on
$\alpha_1$ by relying only on an unambiguously present velocity such
as ${\bf v}_\oplus$.

It is straightforward to generalize the results of the previous
section to the case of a velocity ${\bf v}_\oplus$ which is not
constant, as opposed to $\bf w$. The equation satisfied by the
eccentricity vector $\bf e$ is Eq. (\ref{eq:27}), but
we concentrate now on the effects generated by the source term
$\bbox{\kappa}_\oplus\times {\bf c}$, where $\bbox{\kappa}_\oplus$,
given by Eq.  (\ref{eq:19bis}), is a vector rotating with angular
frequency $n_\oplus \equiv (Gm_\oplus/r_{\oplus\odot}^3)^{1/2}$. As
in the previous section, it suffices to decompose
$\bbox{\kappa}_\oplus\times{\bf c}$ as a sum of constant-norm vectors
rotating with constant angular frequencies to derive the
contributions to $\bf e$ involving
$v_\oplus$. Let us denote as  $I_\oplus=
23.5^\circ$ the inclination of the ecliptic with respect to the
Earth's equatorial plane, and choose the origin of time at the vernal
equinox. The source term $\bbox{\kappa}_\oplus\times{\bf c}$ can then
easily be decomposed as
\begin{equation}
\bbox{\kappa}_\oplus\times{\bf c} = -\alpha_1\ {Gm_\oplus\over 4 a^2
c^2}\ v_\oplus\, ({\bf K}_{++} +{\bf K}_{+-} +{\bf K}_{-+} +{\bf
K}_{--} + {\bf K}_{0+} +{\bf K}_{0-})
\ ,
\label{eq:46}
\end{equation}
where
\begin{equation}
{\bf K}_{0\pm}\equiv-{1\over 2}\ \sin I\,\sin I_\oplus\,
\bigl[\cos(n_\oplus t)\ {\bf a}\mp \sin(n_\oplus t)\ {\bf
b}\bigr]
\ , \label{eq:47}
\end{equation}
and if $s$, $s'$ denote two independent signs ($s = \pm 1$, $s'=\pm
1$)
\begin{equation}
{\bf K}_{ss'} \equiv
s'\ {1+s\cos I\over 2}\ {1-ss'\cos I_\oplus\over 2}\ 
\bigl[
\cos(s\Omega+s'n_\oplus t)\ {\bf a}
-\sin(s\Omega+s'n_\oplus t)\ {\bf b}
\bigr]
\ .
\label{eq:48}
\end{equation}
The vectors ${\bf K}_{0\pm}$ are rotating with angular velocities
$\mp n_\oplus$ in the orbital plane, whereas ${\bf K}_{ss'}$ are
rotating with angular velocities $-(s\dot\Omega+s'n_\oplus)$ in this
plane. These notations are chosen to simplify the
expression of the eccentricity vector $\bf e$ written below; it
generalizes the notations ${\bf K}_\pm$,${\bf K}_0$ introduced in
section IV, which would be denoted ${\bf K}_{\pm 0}$,${\bf
K}_{00}$ in the present convention. The eccentricity vector can
then be immediately written as
\begin{eqnarray}
{\bf e} = {\bf e}_N + \alpha_1\ {Gm_\oplus\over 4 a^2 c^2}\
v_\oplus &&\biggl(
{{\bf K}_{++}\over\dot\omega_N+\dot\Omega+n_\oplus} +
{{\bf K}_{+-}\over\dot\omega_N+\dot\Omega-n_\oplus}
\nonumber\\
&&+{{\bf K}_{-+}\over\dot\omega_N-\dot\Omega+n_\oplus} +
{{\bf K}_{--}\over\dot\omega_N-\dot\Omega-n_\oplus} +
{{\bf K}_{0+}\over\dot\omega_N+n_\oplus} +
{{\bf K}_{0-}\over\dot\omega_N-n_\oplus}
\biggr)
\ ,
\label{eq:49}
\end{eqnarray}
to which should be added the contributions proportional to $w$ of
Eq. (\ref{eq:32}). The solution (\ref{eq:49}) now exhibits
resonances for any value of the inclination $I$ provided the
semi-major axis $a$ is chosen appropriately. Figure 3 displays the
values of $a$ and $I$ for which one of the divisors in Eq.
(\ref{eq:49}) vanishes. Note that when $a$ tends formally
towards $0$, these curves tend towards the six poles found
in section IV, because
$\dot\omega_N$ and $\dot\Omega$ [Eqs. (\ref{eq:28}) and
(\ref{eq:26})] are proportional to $a^{-7/2}$ whereas
$n_\oplus=2\pi/(1\ {\rm yr})$ is constant.

As in the previous section, the price to pay for a small divisor is
the need for a correspondingly long time of observation,
say $T/4=\pi/2(\dot\omega_N\pm\dot\Omega\pm n_\oplus)$, or
$\pi/2(\dot\omega_N\pm n_\oplus)$, depending upon the concerned
divisor. The figure of merit $\delta\rho/(T/4)$ of the different
orbits can be computed like in Eqs. (\ref{eq:37}), and Figure 4
shows that the contributions ${\bf e}_{+-}$ and ${\bf e}_{-+}$ give
the best results, {\it i.e.} the largest perigee displacements in a
given observational time. This is due to the rather small value of
the inclination $I_\oplus=23.5^\circ$ of the ecliptic with respect
to the equatorial plane, since ${\bf e}_{+-}$ and ${\bf e}_{-+}$ are
precisely the only contributions which do not vanish as
$I_\oplus\rightarrow 0$. The values of $a$ and $I$ for which these
two main contributions diverge have been plotted in solid lines in
Fig. 3 in order to distinguish them from those involving $\sin
I_\oplus$ or $1-\cos I_\oplus$ (vanishing when $I_\oplus\rightarrow
0$) which have been plotted in dashed lines.

The largest figure of merit displayed in Fig. 4 is obtained for
$a\approx R_\oplus$ and $I=43.22^\circ$, and reads
$4\delta\rho_{+-}/T_{+-} = \alpha_1\times 8.62\times 10^3\ {\rm
cm.yr}^{-1}$. As expected, it is ($w/v_\oplus\approx$) 12 times
smaller than the best result (\ref{eq:37a}) of section IV, which was
obtained for $I=46.38^\circ$ (or $133.62^\circ$). However, the
high-precision observation of a low orbit with an inclination close
to $43.22^\circ$ could lead to the first experimental results about
preferred-frame effects free from any assumption about the
fundamental rest frame. [The satellite LACE happens to be precisely
on such an orbit, but it is not drag-free.]

We have seen in section IV that all of the three contributions ${\bf
e}_+$, ${\bf e}_-$ and ${\bf e}_0$ (proportional to $w$) to the
eccentricity vector $\bf e$ are {\it a priori\/} large for the orbit
of LAGEOS I, thus giving rise to a complex signal in the evolution
of the eccentricity vector. It is interesting to quote that the
contribution ${\bf e}_{-+}$ (proportional to $v_\oplus$) is also
enhanced by a small divisor for this satellite, since the amplitude
$\delta\rho_{-+}$ is about $\alpha_1\times 2\times 10^3\ {\rm cm}$ in
a typical observational time of $T_{-+}/4\sim 7$ months. It would be
therefore interesting to analyze the LAGEOS I data to look for
preferred-frame signals of both the $\bf w$ and the ${\bf
v}_\oplus$ types [if it turns out feasible to extract such
information in presence of the non-modelizable forces that act on
the satellite].

A remarkable feature of Fig. 3 is the existence of maxima in the
curves representing the loci of the ${\bf v}_\oplus$-type resonances
in the $(I,a)$ plane. These maxima occur at $I=0$ or $180^\circ$ (for
$a/R_\oplus = 1.94$, $2.36$ and $2.65$), $I= 78.46^\circ$ or
$101.54^\circ$ (for $a/R_\oplus = 1.67$), and $I= 90^\circ$ (for
$a/R_\oplus = 1.59$). The amplitude of the ${\bf
v}_\oplus$-preferred-frame effects for orbits
located close to these values of $I$ and $a$ would be therefore
almost insensitive to injection errors or fluctuations in $I$. This
suggests that these orbits might be especially robust tools for
constraining this type of effects.
The best choices would be the orbits located close to ($I=0$, $a =
1.94 R_\oplus$) or ($I= 101.54^\circ$,  $a=1.67 R_\oplus$), which
correspond to rather large figures of merit
$4\delta\rho/\alpha_1 T$ in Fig. 4 (respectively $5.15\times 10^3\
{\rm cm.yr}^{-1}$ and $3.58\times 10^3\ {\rm cm.yr}^{-1}$).

\subsection{Along-track perturbations of the satellite}
We have not yet discussed the perturbation of the element $\sigma$,
which is related to the angular position of the satellite at
$t=0$. Using the expression of the disturbing
function (\ref{eq:10}) in (\ref{eq:9d}), the secular evolution of
$\sigma$ is given by
\begin{equation}
\left\langle{d\sigma\over dt}\right\rangle
= \dot\sigma_N +{\alpha_1\over 4}\ {n^2 a\over c^2}\ {[{\bf w}+{\bf
v}_\oplus, {\bf c}, {\bf e}]\over e^2} - {\alpha_1 n\over c^2}\,
({\bf w}+{\bf v}_\oplus)^2 +O(\alpha_1 e)
\ ,
\label{eq:50}
\end{equation}
where $\dot\sigma_N$ is the Newtonian contribution due principally
to tidal forces (\ref{eq:14}) and to the Earth's quadrupolar moment
(\ref{eq:11}). The second term of the right-hand-side, though
proportional to $1/e$, does not lead to any interesting
effect in the limit of a small eccentricity. Indeed, it
precisely cancels a corresponding term in the secular evolution of
$\omega$. In other words, the angular position of the satellite with
respect to the ascending node is not affected by this
type of contribution.

By contrast, the last term proportional to $({\bf w}+{\bf
v}_\oplus)^2={\bf w}^2+{\bf v}_\oplus^2+2({\bf w}\cdot{\bf
v}_\oplus)$ induces interesting effects on the position of the
satellite through the time variation of ${\bf w}\cdot{\bf v}_\oplus$.
[The secular drift in $\sigma$ due to the constant term
${\bf w}^2+{\bf v}_\oplus^2$ is unobservable because, as we shall
see below, it can be absorbed in a small renormalization of Newton's
constant $G$.] Indeed, let us denote
as before the right ascension and declination of $\bf w$ as
$(\alpha, \delta)$, the orbital frequency of the Earth as
$n_\oplus$, and the inclination of the ecliptic
as $I_\oplus$. Equation (\ref{eq:50}) can then
be rewritten as
\begin{equation}
\langle\dot\sigma\rangle = {\rm cst} - 2 \alpha_1 n\, {w
v_\oplus\over c^2} \left[{1+\cos I_\oplus\over 2} \cos\delta\,
\sin(n_\oplus t-\alpha) +O(1-\cos I_\oplus)+O(\sin
I_\oplus)\right]
\ ,
\label{eq:51}
\end{equation}
where we have written down only the largest contribution. The
integration of this equation shows that the longitude of the
satellite is modulated by an oscillating term
\begin{equation}
\delta\sigma = \alpha_1\ {n\over n_\oplus}\ {w\over c}\
{v_\oplus\over c}\, (1+\cos I_\oplus)\, \cos\delta\,
\cos(n_\oplus t-\alpha)
\ ,
\label{eq:52}
\end{equation}
corresponding to an along-track oscillation
\begin{equation}
\delta x_{/\!/} = a\,\delta\sigma
\approx \alpha_1(R_\oplus/a)^{1/2}\cos(n_\oplus t-\alpha)\times
9.17\times 10^5\ {\rm cm.}
\label{eq:53}
\end{equation}
The large amplitude, yearly periodicity, and weak altitude
dependence ($\propto a^{-1/2}$) of this effect makes it a promising
way of improving the precision of measurement of $\alpha_1$ (maybe
down to the $10^{-6}$ level). However, detailed numerical
simulations are needed to assert whether the effect (\ref{eq:53})
can be separated from the other yearly perturbations.

Let us note finally that the results (\ref{eq:52}),(\ref{eq:53}) can
be derived in a totally different way, by starting directly from the
Lagrangians (\ref{eq:1}) and (\ref{eq:8}). The terms proportional to
$\alpha_1$ can indeed be interpreted as a time-dependent
renormalization of the gravitational constant
\cite{W71,NW72} experienced by the satellite
\begin{equation}
G(t) \equiv G\left[1-{\alpha_1\over 2c^2}({\bf w}+{\bf v}_\oplus)^2
\right]
\ .
\label{eq:54}
\end{equation}
Taking into account the adiabatic constancy of the Delaunay variable
(or ``action variable") $L\equiv [G(t)m_\oplus a(t)]^{1/2}$, Eq.
(\ref{eq:54}) induces a variation of the semi-major axis
\begin{equation}
a(t) \equiv a_N\left[1-{\alpha_1\over 2c^2}({\bf w}+{\bf v}_\oplus)^2
\right]^{-1}
\ .
\label{eq:55}
\end{equation}
In the limiting case of small eccentricity, the angular position
$\theta$ of the satellite with respect to the perigee can be
identified with the mean anomaly $\ell = \int n(t)\,dt + \sigma_N$,
where $\sigma_N$ is a constant and where the orbital frequency $n(t)$
reads
\begin{equation}
n(t) = \left({G(t) m_\oplus\over a(t)^3}\right)^{1/2} =
n_N\left[1-{\alpha_1\over 2c^2}\,({\bf w}+{\bf v}_\oplus)^2 \right]^2
\ .
\label{eq:56}
\end{equation}
Integrating Eq. (\ref{eq:56}) reproduces the result (\ref{eq:52}).
Note that here again a very small instantaneous perturbation
$O(v^2/c^2)$ has been enhanced by a large factor $n/n_\oplus$.
As an aside, let us remark that another consequence of the
time-dependent renormalization (\ref{eq:54}) of the gravitational
constant is to cause a yearly ``breathing" of the radius of the
Earth, with associated yearly variations of its moment of inertia
and of its angular velocity (see \cite{W81}). However, the amplitude
of these variations, given the present limits on $\alpha_1$, are too
small to be of observational significance, {\it e.g.} $\delta
R_\oplus(t) = -(\partial\ln R_\oplus/\partial\ln
G)\,\alpha_1 R_\oplus\,{\bf w}\cdot{\bf v}_\oplus(t)/c^2 < 0.4\ {\rm
mm}$.

\section{Conclusions}
Artificial Earth satellites can be very useful tools to probe the
field content of gravity, {\it i.e.} specifically whether it
contains a vector or second tensor interaction leading to
preferred-frame effects in local gravitating systems. Thanks to the
appearance of small divisors which enhance the preferred-frame
effects on the eccentricity vector when the inclination and/or the
semi-major axis of an orbit are chosen appropriately, it seems
conceivable to tighten the present experimental bounds on
the preferred-frame parameter $\alpha_1$ down to the
$10^{-5}$---$10^{-6}$ level. What is needed are centimeter-level
tracking data of a (naturally or artificially) drag-free satellite
over time scales large enough to separate from Newtonian
contributions the secular motion of the eccentricity vector induced
by a non-zero $\alpha_1$. Among the class of zero-inclination
(equatorial) orbits, geostationary ones are nearly optimal.
Along-track oscillations of a satellite with yearly period constitute
another promising way of measuring $\alpha_1$ around the $10^{-6}$
level. There is also a wide class of orbits for which preferred-frame
effects on the eccentricity vector due to the orbital velocity of the
Earth around the Sun are enhanced by small divisors. These could be
used to obtain the first bounds on $\alpha_1$ independent of any
hypothesis concerning the gravitationally preferred rest frame.

In this paper, we have presented an approximate analytical treatment
of the dominant preferred-frame effects. The value of this treatment
is mainly indicative, as a help for selecting the most favorable
orbits. In practice, we advise to resort to direct numerical
integration of the equations of motion (and, evidently, to a
multiparameter fit to the experimental data). For the convenience of
the interested reader, we end by giving the $\alpha_1$ (and for
completeness $\alpha_2$) contributions to the equations of motion.
Contrary to the rest of the paper, we have in mind here global
(barycentric) equations of motion (written in a post-Newtonian
coordinate system appropriate to the description of the entire solar
system; see {\it e.g.} \cite{DSX4}). The relative acceleration of a
satellite with respect to the Earth has the form
\begin{equation}
{d^2{\bf x}\over dt^2} = {\bf A}_{\rm GR} +{\bf A}_{\rm
non\hbox{-}grav}+{\bf A}_{\alpha_1} + {\bf A}_{\alpha_2}
\ ,
\label{eq:57}
\end{equation}
where the general relativistic geocentric acceleration (including a
relativistic treatment of multipolar and tidal effects) will be
found in full detail in Ref. \cite{DSX4}, where ${\bf A}_{\rm
non\hbox{-}grav}\approx {\bf F}_{\rm non\hbox{-}grav}/m_{\rm sat}$
denotes the acceleration induced by non-gravitational forces, and
where (to lowest order in the deviation from general relativity, and
for $m_{\rm sat}\ll m_\oplus$, $r\ll r_{\oplus\odot}$)
\begin{mathletters}
\label{eq:58}
\begin{eqnarray}
{\bf A}_{\alpha_1} = &&\alpha_1\,{Gm_\oplus\over 2r^2c^2}\left\{
w_\oplus^2\,{\bf n}-({\bf n}\times{\bf
w}_\oplus)\times{\bf v}\right\}\nonumber\\
&&+\alpha_1\,{Gm_\odot\over
2r_{\oplus\odot}^2c^2}\left\{({\bf n}_{\oplus\odot}\times{\bf
w})\times{\bf v}+\left(2\left[{E^{\rm grav}_\oplus\over
m_\oplus}\right]+{Gm_\oplus\over r}\right){\bf
n}_{\oplus\odot}\right\}
\ ,
\label{eq:58a}\\
{\bf A}_{\alpha_2} = &&-\alpha_2\,{Gm_\oplus\over 2r^2c^2}\left\{
({\bf n}\times{\bf w}_\oplus)^2\,{\bf n}+2({\bf n}\cdot{\bf
w}_\oplus)\,({\bf n}\times{\bf w}_\oplus)\times{\bf n}\right\}
\nonumber\\
&&-\alpha_2\,{Gm_\odot\over
2r_{\oplus\odot}^2c^2}\left\{{4\over 3}\,\left[{E^{\rm
grav}_\oplus\over m_\oplus}\right]{\bf
n}_{\oplus\odot}+{Gm_\oplus\over r}\,({\bf n}\times{\bf
n}_{\oplus\odot})\times{\bf n}\right\}
\ .
\label{eq:58b}
\end{eqnarray}
\end{mathletters}
In these formulae, $r$ and $\bf v$ denote the radial distance
and the velocity of the satellite with respect to the Earth, $\bf n$
the unitary vector directed from the Earth to the satellite, ${\bf
n}_{\oplus\odot}$ the unitary vector directed from the Earth to the
Sun\cite{F5}, and ${\bf w}_\oplus = {\bf w} + {\bf v}_\oplus$ the
absolute velocity of the Earth with respect to the gravitationally
preferred rest frame, ${\bf v}_\oplus$ being its orbital velocity
around the Sun. For conceptual clarity we have indicated within
square brackets the contributions of $\alpha_1$ and $\alpha_2$ due to
the violation of the strong equivalence principle, {\it i.e.} the
terms generated by expanding the $\eta$-dependent term coming
{}from the Lagrangian $L_{\beta,\gamma,\eta}$, Eq. (\ref{eq:2}). The
different terms of Eqs. (\ref{eq:58}) are classified by order of
decreasing magnitude. [The even smaller contributions due to the
coupling of the satellite to the Earth's intrinsic angular momentum
are given in Eq. (9.20) of Ref. \cite{W81}; they cause an
additional secular precession of the satellite's orbit which is
independent of the absolute velocity with respect to the preferred
frame.]

The contribution proportional to $w_\oplus^2\,{\bf n}$ in
Eq. (\ref{eq:58a}) is responsible for the along-track perturbations
of the satellite studied in section V.B above, whereas the one
proportional to $({\bf n}\times{\bf w}_\oplus)\times{\bf v}$,
together with the contribution proportional to $(Gm_\oplus/r)\,{\bf
n}_{\oplus\odot}$ in the second line, is responsible for the perigee
displacements studied in the rest of the paper. Note that the
$(Gm_\oplus/r)\,{\bf n}_{\oplus\odot}$ term has the same form as
the one due to a violation of the strong equivalence principle
[with an opposite sign: $2E^{\rm grav}_\oplus/m_\oplus c^2\approx
-9.2\times 10^{-10}$, $Gm_\oplus/a c^2\approx
(R_\oplus/a)\times 7.0\times 10^{-10}$]. Actually, neither of these
two terms is of much observational significance, as their
contributions are much smaller than that due to the $({\bf
n}\times{\bf w}_\oplus)\times{\bf v}$ term.

The force proportional to $({\bf n}_{\oplus\odot}\times{\bf
w})\times{\bf v}$ in the second line of Eq. (\ref{eq:58a}) has not
been considered in the rest of the paper. It arises in the
connection between the locally inertial geocentric frame used in the
body of the paper and the global barycentric one used here, and has
the form of a Coriolis force. This Coriolis force acts also on
gyroscopes (including the spinning Earth) and adds up to several other
relativistic effects causing a universal precession $\bf\Omega$ of
gyroscopes and satellite's orbits with respect to the barycentric
frame (see section 9.1 of \cite{W81} and \cite{DSX4}). The
$\alpha_1$-dependent contribution to $\bf\Omega$ reads
\begin{equation}
{\bf\Omega}_{\alpha_1} = \alpha_1\ {Gm_\odot\over
4r_{\oplus\odot}^2c^2}\ {\bf n}_{\oplus\odot}\times{\bf w}
\ .
\label{eq:59}
\end{equation}
Although the time dependence of ${\bf\Omega}_{\alpha_1}$, {\it via\/}
${\bf n}_{\oplus\odot}(t)$, might in principle allow one to separate
it from the other relativistic contributions to $\bf\Omega$ (when
discussing observables related to the global barycentric frame), its
magnitude is too small to be of observational significance. [It
yields a yearly oscillation of the satellite with respect to the
barycentric frame which is smaller than $(a/R_\oplus)\times 0.1\ {\rm
mm}$.]


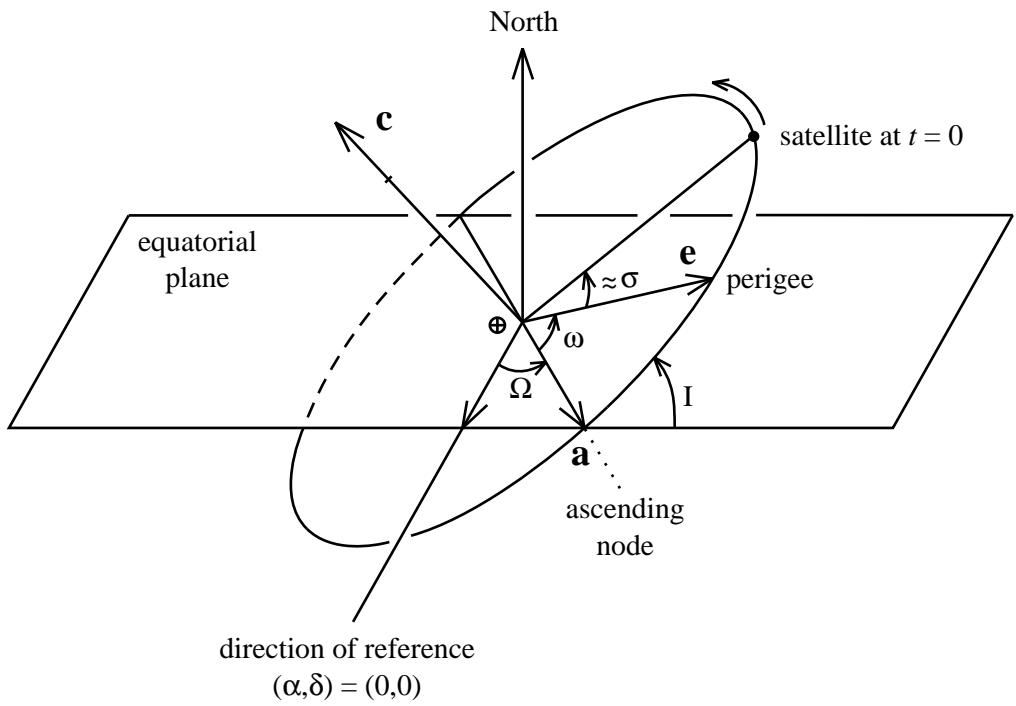
\begin{figure}
\caption{Definition of the orbital elements $\bf e$, $I$, $\Omega$,
$\omega$, and of the rotating orthonormal basis $({\bf a}, {\bf b}
= {\bf c}\times{\bf a}, {\bf c})$ linked with the orbital plane. In
the limit of a small eccentricity $e$, the angle between the
perigee and the position of the satellite at $t=0$
can be identified with the orbital element $\sigma$.}
\end{figure}

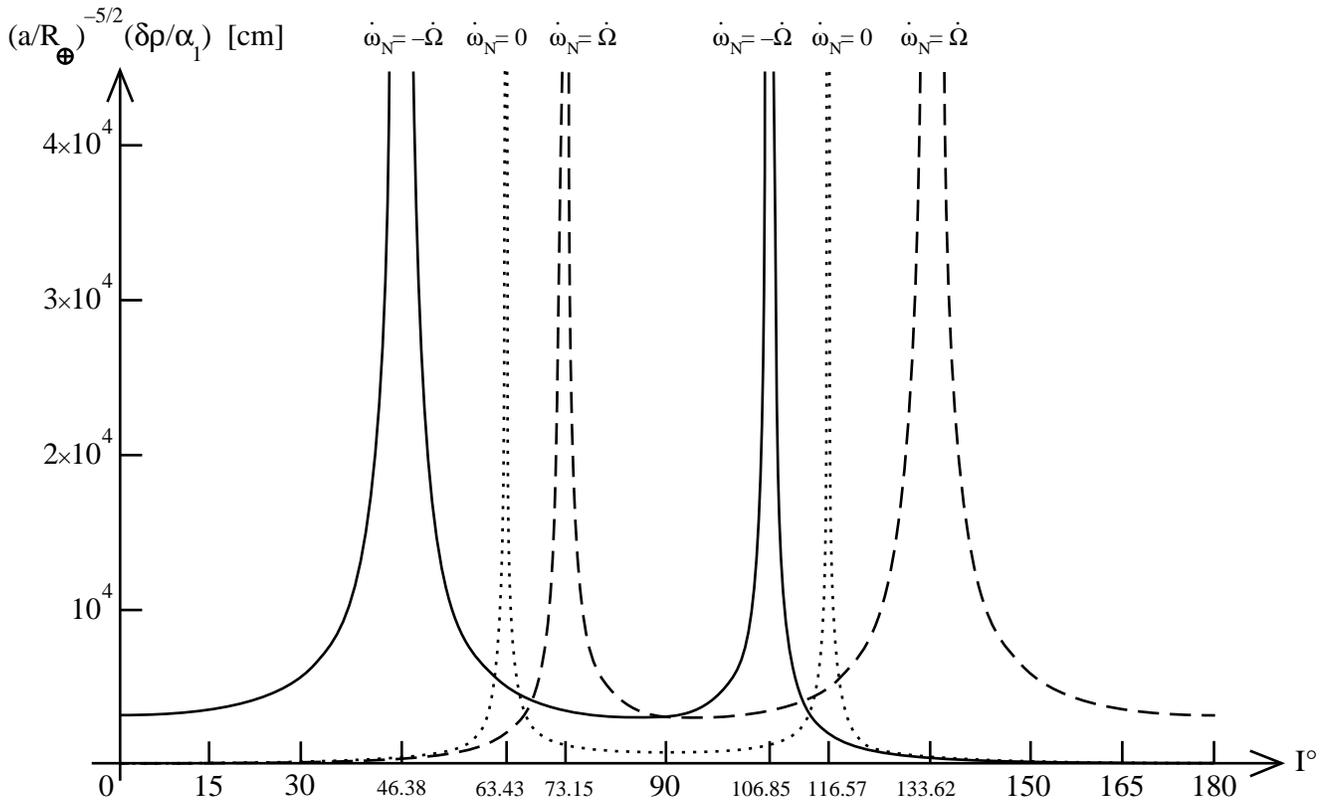
\begin{figure}
\caption{Expected radial displacements of the perigee as functions of
the inclination of the orbital plane. The solid line correspond to
the contribution $\delta\rho_+$, the dashed line to $\delta\rho_-$,
and the dotted line to $\delta\rho_0$.}
\end{figure}

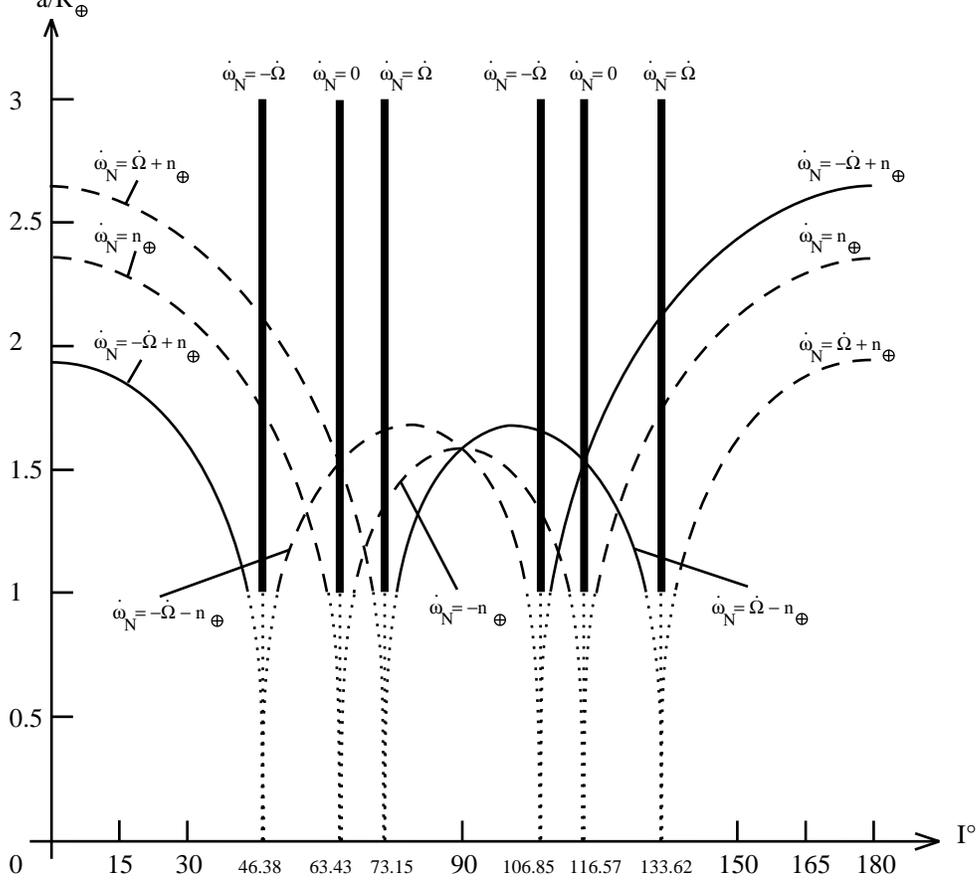
\begin{figure}
\caption{Values of the semi-major axis of the satellite's orbit for
which the displacement $\delta\rho$ of the perigee has a resonance, as
functions of the inclination of the orbital plane. The bold lines
correspond to the six poles shown on Fig. 2, which exist for any
value of $a$ (small enough for tidal forces to be negligible). The
solid lines correspond to effects proportional to $v_\oplus$ which do
not vanish when the inclination $I_\oplus$ of the ecliptic with
respect to the equatorial plane is neglected. The dashed lines
correspond to the vanishing effects when $I_\oplus\rightarrow 0$.
The dotted lines correspond to unphysical values of $a<R_\oplus$.}
\end{figure}

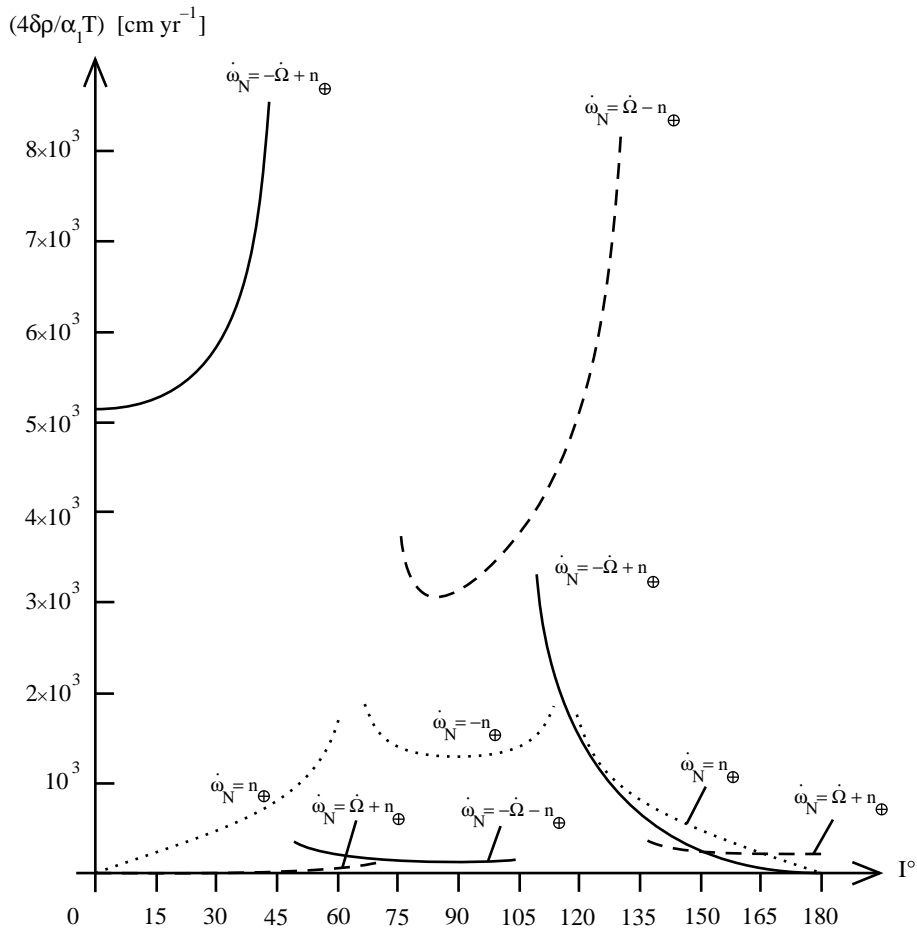
\begin{figure}
\caption{Figure of merit of the resonant preferred-frame effects
corresponding to the orbits of Fig. 3, {\it i.e.} ratio of the
expected radial displacement $\delta\rho$ of the perigee by the
typical observational time $T/4$. The plain lines correspond to the
contribution $\delta\rho_{+\pm}$, the dashed lines to
$\delta\rho_{-\pm}$, and the dotted lines to $\delta\rho_{0\pm}$.}
\end{figure}


\begin{references}
\bibitem{F1}In the present paper, we shall limit our consideration
to the four post-Newtonian parameters $\beta$, $\gamma$, $\alpha_1$,
$\alpha_2$, with special attention to $\alpha_1$. Although extended
versions of the parametrized post-Newtonian (PPN) formalism have
been defined, containing up to ten PPN parameters, the extra six
parameters seem to be less well-rooted in plausible alternative
field theories of gravitation.
\bibitem{F2}We follow the standard astronomical conventions in
denoting by $\delta$ the declination (angle above the Earth equator;
$\delta = +\pi/2$ for the North pole) and by $\alpha$ the right
ascension (longitude around the Earth polar axis counted from the
vernal point).
\bibitem{F3}Within the present approach, the center of mass is
defined using the standard Newtonian formulas in the preferred frame.
\bibitem{F4}To simplify our rough analytical estimates we assume
that the $\sqrt{N}$ improvement due to the number of individual
measurements just compensates the difficulty of determining the
center of a circle from a limited segment ($\sim$ a quarter).
\end{references}
\end{document}